\documentclass[twocolumn,english,aps,prb,showpacs,superscriptaddress]{revtex4-1}
\usepackage[T1]{fontenc}
\usepackage[latin9]{inputenc}
\usepackage{textcomp}
\usepackage{amsmath}
\usepackage{amssymb}
\usepackage{graphicx}
\usepackage{color}

\makeatletter

\usepackage{ulem}
\usepackage{float}
\usepackage{lipsum}
\usepackage{babel}

\makeatother

\usepackage{babel}
\begin{document}

\title{Localization, multifractality, and many-body localization in periodically kicked quasiperiodic lattices}

\author{Yu Zhang}
\affiliation{Beijing National Laboratory for Condensed Matter Physics, Institute
of Physics, Chinese Academy of Sciences, Beijing 100190, China}
\affiliation{School of Physical Sciences, University of Chinese Academy of Sciences,
Beijing 100049, China }
\author{Bozhen Zhou}
\affiliation{Beijing National Laboratory for Condensed Matter Physics, Institute
of Physics, Chinese Academy of Sciences, Beijing 100190, China}
\affiliation{School of Physical Sciences, University of Chinese Academy of Sciences,
Beijing 100049, China }
\author{Haiping Hu}
\affiliation{Beijing National Laboratory for Condensed Matter Physics, Institute
of Physics, Chinese Academy of Sciences, Beijing 100190, China}

\author{Shu Chen}
\thanks{schen@iphy.ac.cn }

\affiliation{Beijing National Laboratory for Condensed Matter Physics, Institute
of Physics, Chinese Academy of Sciences, Beijing 100190, China}
\affiliation{School of Physical Sciences, University of Chinese Academy of Sciences,
Beijing 100049, China }
\affiliation{Yangtze River Delta Physics Research Center, Liyang, Jiangsu 213300,
China }
\date{\today}

\date{\today}
\begin{abstract}
We study the combined effect of quasiperiodic disorder, driven  and interaction  in the periodically kicked  Aubry-Andr\'{e} model.
In the non-interacting limit, by analyzing the quasienergy spectrum statistics, we verify the existence of a dynamical localization transition in the high-frequency region, whereas the spectrum statistics becomes intricate in the low-frequency region due to the emergence of the extended/localized-to-multifractal edges in the quasienergy spectrum, which separate the multifractal states from the extended (localized) states.  When the interaction is introduced, we find the periodically
kicked incommensurate potential can lead to a transition from ergodic to many-body-localization phase in the high-frequency
region. However, the many-body localization phase vanishes in the low-frequency region even for strong quasiperiodic disorder. Our studies demonstrate that the periodically kicked  Aubry-Andr\'{e} model displays rich dynamical phenomena and the driving frequency plays an important role in the formation of many-body localization in addition to the disorder strength.

\end{abstract}
\keywords{w}
\maketitle

\section{Introduction}
Anderson localization is a fundamental phenomenon of quantum disorder systems
and has attracted longstanding attention in condensed matter physics \citep{Anderson_disorder,Anderson 50 year,Mirlin_anderson_transition}. While localization-delocalization  transition and mobility edges only occur in three dimensions for random disorder systems, the localization-delocalization transition can be found in one-dimensional quasiperiodic systems, which
have attracted increasing interest in recent years \cite{AAmodel,AAmodel2,Bloch2018,Roati,Lucioni}.
When the quasiperiodical potential strength exceeds a critical value, a localization transition takes place as illustrated by the prototypical quasiperiodic model known as the Aubry-Andr\'{e} (AA) model \citep{AAmodel,AAmodel2}.  The quasiperiodic optical lattices have become an ideal platform for studying the localization-delocalization transition \cite{Bloch2018,Roati,Lucioni}.
Particularly, the interplay of interaction and disorder can induce many-body localization (MBL) \citep{MBL1,MBL2,MBL3,David Huse ratio,MBL5}, which violates
eigenstate thermalization hypothesis  and prevents erogdicity \citep{LIOM1,LIOM2}. The existence of MBL has been confirmed in one-dimensional interacting systems with random disorder \citep{David Huse entropy MBL tansition,MBL-random1,MBL-random2,MBL-random3,David Huse ratio,Gray,David Huse entropy MBL tansition,MBL-EE1}
or incommensurate potential\citep{MBL-inc1,MBL-inc2,MBL-inc3,MBL-inc4,MBL-inc5,MBL-inc6,MBL-inc7,MBL-inc9,MBL-inc10,MBL-inc11,MBL-inc12,MBL-inc13,WangYC2021,XuSL,Aramthottil,Sierant}.
Moreover, the MBL phase has been experimentally observed in the ultracold atomic gases trapped in incommensurate optical lattices \citep{MBL-expeiment1,MBL-experiment2,MBL-experiment3,MBL-experiment4,MBL-experiment5}.

Exploring novel non-equilibrium phases in driven, interacting quantum systems is a topic of perennial interest. In general, periodically driving a quantum system results in thermalization of the system \cite{drivingETH1,drivingETH2}. Nevertheless, recent works have demonstrated the existence of MBL which allows avoiding heating in the
presence of driving \citep{F_MBL2,F_MBL1,Abanin Thouless Energy Floquet,Abanin MBL Floquet 2015,Huse Floquet for MBL}.
The combination of MBL and Floquet driving can lead to new non-equilibrium
phases of matter, such as time crystals \citep{TmCr1,TmCr2}, suggesting that the interplay of periodic driving and MBL would give rise to rich dynamical phenomena.
On the other hand, by applying a pulsed incommensurate potential to an optical lattice, a periodically kicked AA model was proposed to exhibit dynamical Anderson transition \cite{Qin}, which is revealed from its dynamical evolution of wave packets. A dynamical localization is characterized
by the halt of the spreading of an initial wave packet, and the transition depends on both the strength of the quasiperiodic potential and the kicking period \cite{Qin,Sacramento,Santhanam}.
For the periodical kicked case, while the
time evolution is governed by an effective time-independent AA model in the high-frequency region, the dynamics in the low-frequency region is far more intricate and has not yet been well understood. Besides the kicked AA model, other periodically driving quasiperiodic models are also studied \cite{Ghosh,Sarkar,YiXX,Sarkar2}, and the effect of temporal disorder on the wave-packet dynamics is also analyzed \cite{TongPQ}. Particularly, a recent experiment has observed non-ergodic and ergodic phases in the driven quasiperiodic many-body system which are separated by a drive-induced delocalization transition \cite{Bordia}.

Motivated by these theoretical and  experimental progresses, we shall study the periodically kicked interacting AA model and investigate the combined effect of quasiperiodic disorder, driven period (frequency), and interaction on the dynamical localization by analyzing the
quasienergy spectrum of Floquet operator and the related dynamical behavior. To understand the interplay of the quasiperiodic potential and kicked period, we first analyze the quasienergy spectrum statistics of the noninteracting kicked AA model, which displays different behaviors in the high-frequency and low-frequency region. In the high-frequency region, the spectrum statistics clearly demonstrates a dynamical localization transition signaled by the abrupt change of the average ratio of adjacent quasienergy gaps. In the low-frequency region, the spectrum statistics becomes intricate due to the emergence of the extended/localized-to-multifractal edges, which separate the multifractal states from the localized (extended) states. The corresponding average ratio of adjacent quasienergy gaps is not an universal value and depends on the ratio of numbers of multifratal states and localized (extended) states.
The multifractal states can be identified by finite-size scaling analyze of the corresponding wavefunctions, and we propose a scheme to extract the average multifractal exponent from the long-time survival probability.
We then investigate the interacting kicked AA model and identify the existence of MBL in the high-frequency region. Through the finite-size analyse, we unveil the occurrence of a transition from the ergodic phase to the MBL phase  when the quasiperiodic potential strength exceeds a critical value. However, in the low-frequency region we find that the MBL phase vanishes  and no MBL occurs even for strong quasiperiodic disorder.

The rest of this paper is structured as follows. In Sec. II, we introduce the model and the method of quasienergy spectrum statistics. In Sec. III, we analyze the quasienergy spectral statistics and carry out multifractal analysis for the noninteracting kicked AA model. We unveil the existence of dynamical localization transition in the high-frequency region and the emergence of extended/localized-to-multifractal edges in the low-frequency region. In Sec. IV, we study the MBL in the interacting kicked AA model in detail. A summary is given in the final section.

\section{Model and method}
We consider a periodically kicked quasiperiodic model described by the Hamiltonian
\begin{equation}
H= H_0 + H_K,   \label{H}
\end{equation}
with
\begin{eqnarray}
H_0 &=& H_J + H_V  \nonumber \\
 &=& -J\sum_{j} \left(\hat{c}_{j}^{\dagger}\hat{c}_{j+1}+\text{H.c.}\right)+ \sum_{j}V\hat{n}_{j}\hat{n}_{j+1}
\end{eqnarray}
and
\begin{equation}
H_K=\sum_{n}\delta(t-nT) \sum_{j} \mu_{j}\hat{n}_{j} \label{eq:Hamiltonian},
\end{equation}
where $\hat{c}_{j}^{\dagger}(\hat{c}_{j})$ is the fermion creation
(annihilation) operator, $\hat{n}_{j}=\hat{c}_{j}^{\dagger}\hat{c}_{j}$
is the particle number operator, $J$ is the hopping amplitude between
nearest-neighbor sites, and $V$ is the interaction strength between the neighboring particles. The kicking part of the Hamiltonian is described by $H_K$ with the quasiperiodic potential
\begin{equation}
\mu_{j} = \lambda\cos\left(2\pi j\alpha+\phi\right)
\end{equation}
being periodically added with a pulsed period $T$, where $\alpha=\frac{\sqrt{5}-1}{2}$, $\lambda$ is the strength of the quasiperiodic potential and $\phi$ is a random phase. Taking sample average for the random phase $\phi$  can reduce statistical and finite-size effects.
For convenience we set $\hbar=1$ and take $J=1$ as the unit of energy in the following calculation.
Our model is similar to the interacting spinless fermions model in a quasiperiodic lattice, which has been applied to study MBL \citep{MBL-inc13,MBL-inc4}, but with a periodic kicked potential. We shall demonstrate that the periodically kicked quasiperiodic lattice displays rich dynamical phenomena, including the emergence of multifractal states with no equilibrium counterpart, and the driving frequency plays an important role in the formation of MBL in addition to the quasiperiodic potential strength.

The dynamical evolution of the periodically kicked system
is determined by the Floquet unitary propagator, which can be written as
\begin{align}
U(T)  =e^{-i H_0 T}e^{-i\lambda\sum_{j}^{L}\cos\left(2\pi j\alpha+\phi\right)\hat{n}_{j}} . \label{eq:floquet propagator}
\end{align}
For a given initial
state $\psi(t)$, the finial state after $N$ periods can be written
as $\psi\left(t+NT\right)=\left[U\left(T\right)\right]^{N}\psi\left(t\right)$.
For a Floquet unitary propagator, all the quasienergies are distributed
on the unit circle and we use angles $\theta_{n}$ to denote different quasienergies:
\begin{equation}
\Theta=\left\{ \theta_{n}|\lambda_{n}=e^{i\theta_{n}},\theta_{n}\in[-\pi,\pi)\right\} , \label{eq:theta}
\end{equation}
where $\lambda_{n}$ are the eigenvalues of the operator $U\left(T\right)$ and
$\theta_{n}<\theta_{n+1}$.
In analogy to Hamiltonian systems, we define $s_{n}=\theta_{n+1}-\theta_{n}$, and the level spacing distribution of $\theta$
can be captured by the ratio between adjacent gaps \cite{drivingETH1,Abanin MBL Floquet 2015}:
\begin{equation}
 r_n = \frac{\min\{s_{n},s_{n+1}\}}{\max\{s_{n},s_{n+1}\}} . \label{eq:r}
\end{equation}
The average of $r_n$ is introduced as
\begin{equation}
\langle r\rangle=\frac{1}{\mathcal{D}}\sum_{n=1}^{\mathcal{D}}  r_n , \label{eq:r}
\end{equation}
where $\mathcal{D}=\mathcal{N}-1 $ with $\mathcal{N}$ being the size of Hilbert space.
For the static Hamiltonian system with eigenvalues $E_n$, we have $s_{n}=E_{n+1}-E_{n}$ and the ratio can serve
as a probe of the phase transition between the ergodic and MBL phase \citep{David Huse ratio,MBL5}.
In the ergodic phase the energy level spacings satisfy the Wigner-Dyson
distribution with  $\langle r\rangle \approx 0.529$, whereas in the localized
phase with the Poisson distribution $\langle r\rangle \approx 0.387$ \citep{David Huse ratio}.
This quantity was also applied to study the disordered Floquet systems \cite{meng cheng energy level statistics,Abanin MBL Floquet 2015}.

\section{Spectral statistics and  multifractal analysis  for the kicked AA model}
We first consider the noninteracting case with $V=0$, for which the model (\ref{H}) reduces to the periodically kicked AA model \cite{Qin}.
In the high-frequency region, the
time evolution can be effectively described by the AA model with the critical point given by $\lambda/T =2$. However, in the low-frequency region, the time evolution is far more intricate and has not been fully explored yet \cite{Qin,Sacramento}.

Here we shall scrutinize the kicked AA model by studying the spectral statistics of the quasienergies. To reduce the impact of the edge, we take the lattice size as the Fibonacci
number  and consider the periodic boundary condition (PBC) in the calculation.
In Fig.1(a), we show $\langle r\rangle$ (the average ratio of two consecutive quasienergy gaps) in the parameter space spanned by $\lambda$ and $T$.
In the high-frequency region of $T \ll 1$, it is shown that there is an abrupt transition in $\langle r\rangle$ when the parameter $\lambda$ or $T$ crosses over the diagonal line $\lambda/T=2$. This is also witnessed in Fig.1(b) and 1(c), which indicate an abrupt transition around the diagonal region of  $\lambda/T=2$. When the system is in the dynamically extended region, the
ratio is close to 0, whereas $\langle r\rangle \approx 0.39$ in the region of dynamical localization. It turns out that increasing $\lambda$ can lead to a transition from the extended region to the localized region in the high-frequency region.
On the other hand, when $T>1$, an abrupt transition is observed before reaching the diagonal line of $\lambda/T=2$, due to the emergence of multifractal eigenstates, which are neither fully localized nor fully extended and separated from the extended (localized) eigenstates by extended/localized-to-multifractal edges.
\begin{figure}
\centering{}\includegraphics[scale=0.17]{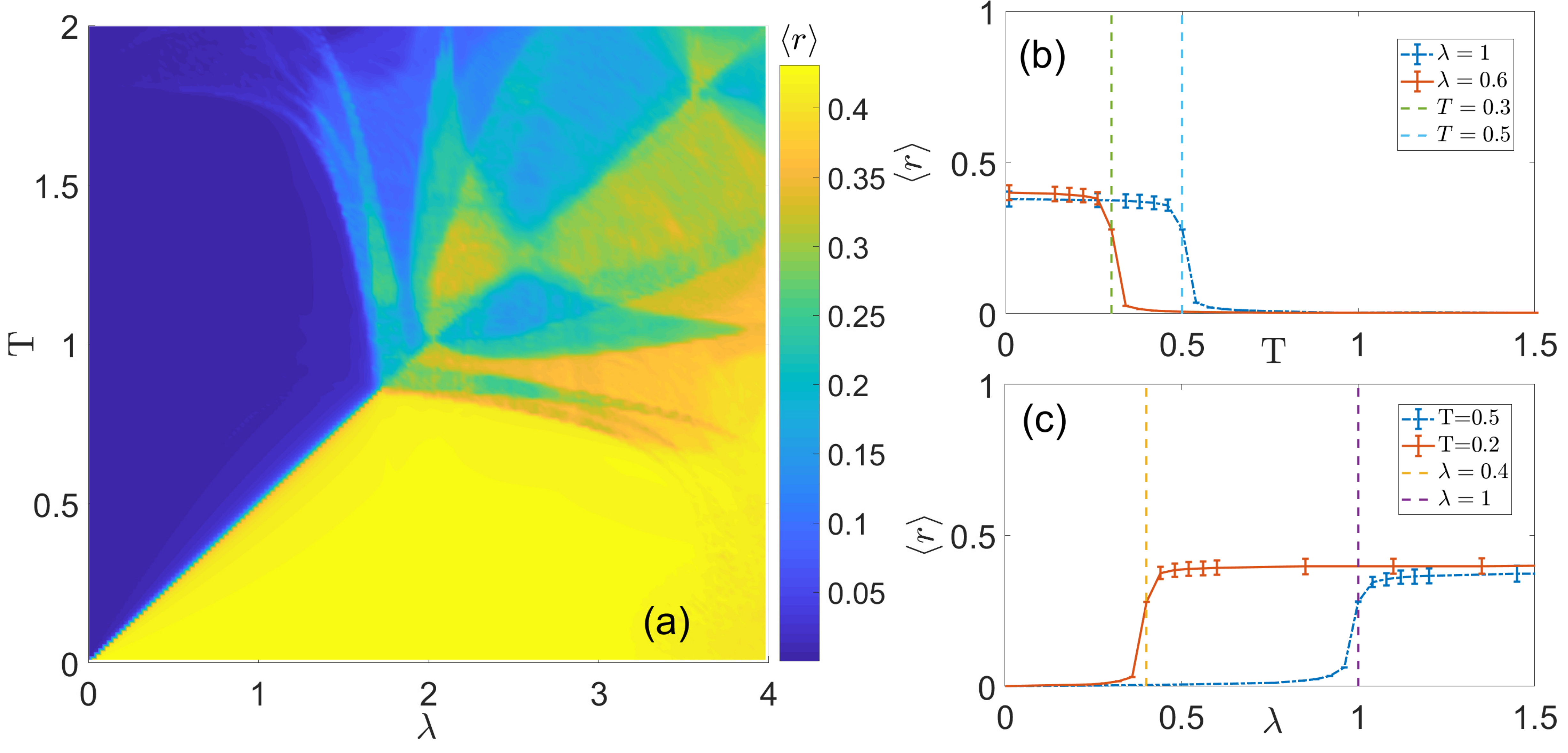}\caption{(a) The average ratio $\langle r\rangle$ in the parameter space spanned by $\lambda$ and $T$ \label{fig:phasediagram}.  (b) $\langle r\rangle$ vs $T$ for fixed $\lambda$.
(c) $\langle r\rangle$ vs $\lambda$ for fixed $T$. The dashed lines are given by $\lambda/T=2$. The system is under the PBC with
$L=987$ and we take 100 samples for each point. \label{fig:r-=00005Clambda}}
\end{figure}

We note that $\langle r\rangle \approx 0$ is due to the existence of nearly double degeneracy in the dynamically extended region.
To see it clearly, we define the even-odd (odd-even) level spacings of the quasienergies as
\[
s_{n}^{e-o}=\theta_{2n}-\theta_{2n-1}\left(s_{n}^{o-e}=\theta_{2n+1}-\theta_{2n}\right).
\]
In Fig.2(a)-(c), we show the even-odd (odd-even)
spacings of the kicked AA model with $L=987$, $T=0.2$ and $\lambda=0.2,0.4,0.6$, corresponding to extended, critical
and localized phases, respectively.
In the extended region, the spectrum is nearly doubly-degenerate and hence there is an obvious  gap between $s_{n}^{e-o}$ and $s_{n}^{o-e}$. In
the localized region, $s_{n}^{e-o}$ and $s_{n}^{o-e}$ have the same
form and the gap vanishes.  In the critical
region, distributions of $s_{n}^{e-o}$ and $s_{n}^{o-e}$ are strongly scattered.
Our results demonstrate that the distribution of quasienergies of the kicked AA model in the high-frequency region displays similar behaviors
as the spectrum distribution of the AA model, for which the even-odd (odd-even)
spacings $s_{n}^{e-o}=E_{2n}-E_{2n-1}$  $\left(s_{n}^{o-e}=E_{2n+1}-E_{2n}\right)$
were utilized to distinguish the different phases of the AA model \citep{xiaolong Deng/PRL}.
\begin{figure}
\begin{centering}
\includegraphics[scale=0.35]{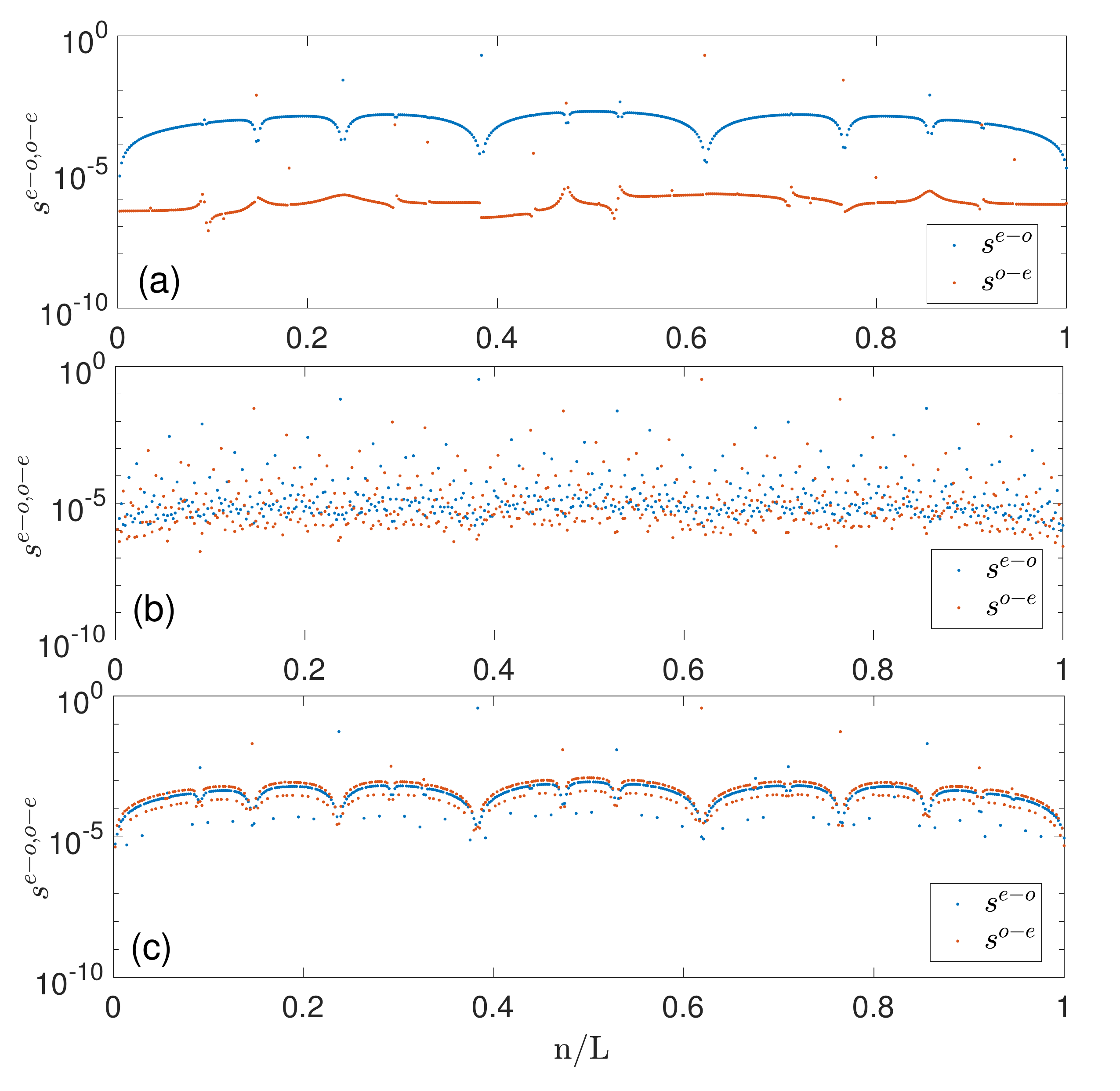}
\par\end{centering}
\centering{}\caption{Even-odd (odd-even) level spacings for the kicked AA model with $T=0.2$ and $L=987$ under the PBC. From top to bottom:
(a) $\lambda=0.2$ (extended), (b) $\lambda=0.4$ (critical), and(c) $\lambda=0.6$ (localized). }
\end{figure}

\begin{figure}
\begin{centering}
\includegraphics[scale=0.29]{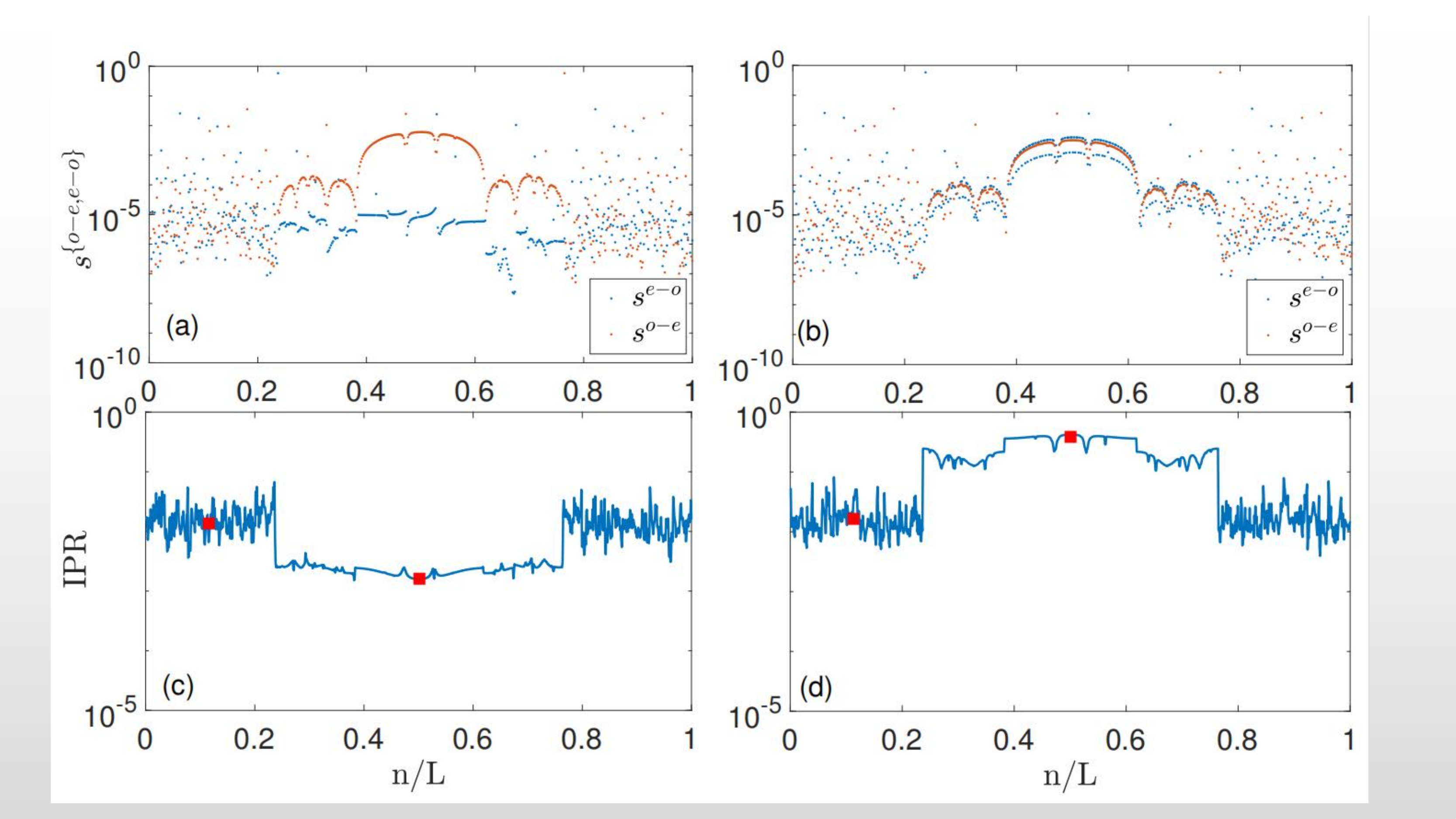} \caption{Even-odd level spacing ($s^{e-o}$) and odd-even level spacing ($s^{o-e}$) for (a) $T=1.5, \lambda=1.6$  and (b) $T=0.8,\lambda=3$, respectively. IPR for (c) $T=1.5,\lambda=1.6$  and (d) $T=0.8,\lambda=3$, respectively. The system is under the PBC with $L=987$.
 \label{fig:mobility-edge}}
\par\end{centering}
\end{figure}

Now we study the low-frequency region where the distributions of $s_{n}^{e-o}$ and $s_{n}^{o-e}$  become intricate. As concrete examples, we consider systems with parameters $\left(\lambda=1.6,T=1.5\right)$
and $\left(\lambda=3,T=0.8\right)$,  which distribute symmetrically about the diagonal line $\lambda=2T$  in the parameter space and can be connected together by a dual transformation (see appendix A) .
The even-odd (odd-even) spacings for these systems are displayed in Fig.\ref{fig:mobility-edge}(a) and (b), respectively. It is shown that the distribution of even-odd (odd-even) spacings exhibits different behavior in the middle and side regions, which are separated by some edges. While there is a gap between $s_{n}^{e-o}$ and $s_{n}^{o-e}$ in the middle region, their distributions are strongly scattered in the the side regions, as shown in Fig.\ref{fig:mobility-edge}(a). The distribution suggests that the states  in the middle and side regions are extended and critical (multifractal) states, respectively. As a contrast, for the system shown in Fig.\ref{fig:mobility-edge}(b), while the distributions in side regions are similar, the gap vanishes in the middle region, suggesting that the states in the middle region are localized states.

To unveil the properties of states more clearly, we also calculate the inverse participation ratio (IPR) for the eigenstate of the
unitary operator, which is defined as $P_n = \sum_{i=1}^{L} |\psi_{n,i}|^4$ with $|\psi_{n} \rangle = \sum_{i=1}^{L} \psi_{n,i} |i \rangle$ representing the $n$-th eigenstate of $U(T)$. As shown in Fig.\ref{fig:mobility-edge}(c) for the system with $\lambda=1.6$ and $T=1.5$, we see $P_n \sim 1/L$ in the middle region, indicating that the corresponding states are extended states. In the side regions, the corresponding states are multifractal (critical) states which are separated from the extended states by the presence of extended-to-multifractal edges.
On the other hand, for the system with $\lambda=3$ and $T=0.8$ as shown in Fig.\ref{fig:mobility-edge}(d), the IPR tends to a finite number in the middle region with the corresponding state being localized. Also there exist localized-to-multifractal edges separating the localized and critical regions. By carrying out a finite-size scaling analysis for the eigenstates, we can distinguish the extended, localized and multifractal states.

\begin{figure}[h]
\begin{centering}
\includegraphics[scale=0.42]{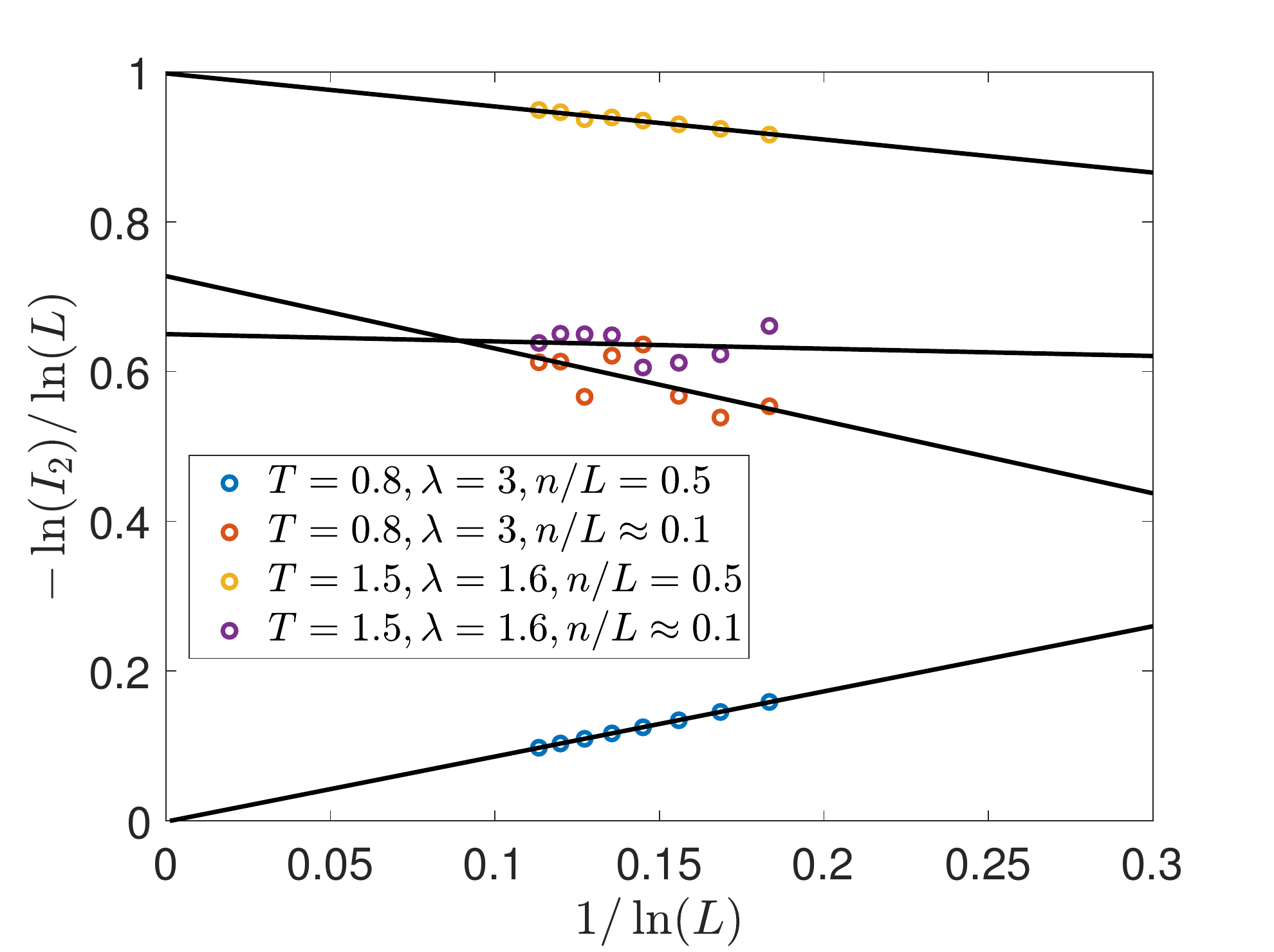}\caption{Finite-size scaling analysis for different eigenstates. \label{fig:multifractal_analysis}}
\par\end{centering}
\end{figure}

Now we carry out a finite-size scaling analysis for eigenstates in
different regions \citep{xiaolong Deng/PRL,Sarkar2,WangYC2016}. For a given eigenstate $|\psi_{n}\rangle=\sum_{j}\psi_{n}(j)|j\rangle$,
we can use the moments
\begin{equation}
I_{q}(n)=\sum_{j}|\psi_{n}(j)|^{2q}\propto L^{-D_{q}(q-1)}
\end{equation}
to characterize the distribution information of the eigenstate. $D_{q}$ are the fractal dimensions
and take difference values in different regions : $D_{q}=1$ in the extended
region, $D_{q}=0$ in the localized region and $0<D_{q}<1$ in the multifractal
region. In our calculation, we choose $q=2$ and the fractal dimensions
can be obtained from the inverse participation ratio $(I_{2})$.
After a simple transformation, it is easy to get
\[
-\ln[I_{2}(n)]/\ln(L)=-c/\ln(L)+D_{2},
\]
where $c$ is a size-independent coefficient. We can get the $D_{2}$
by the intercept of the curve in the space spanned by $1/\ln(L)$ and $-\ln[I_{2}(n)]/\ln(L)$. In Fig.\ref{fig:multifractal_analysis}, we plot the curves
in different regions, as marked by the red squares in Fig.\ref{fig:mobility-edge}(c) and (d). For localized states and extended states, we choose
a typical eigenstate at the middle of the spectrum ($n/L=1/2$). For
multifractal states, we choose $20$ eigenstates near the n-th eigenstate
with $n/L=0.1$ and take an average to eliminate fluctuation.
After a linear fitting, we find that $D_{2}=1$ for extended states
and $D_{2}=0$ for localized states when $L\rightarrow\infty$. We
also find that $0<D_{2}<1$ for the multifractal eigenstates while
$L\rightarrow\infty$. It confirms the existence of the multifractal
states.

\begin{figure}
\centering{}\includegraphics[scale=0.30]{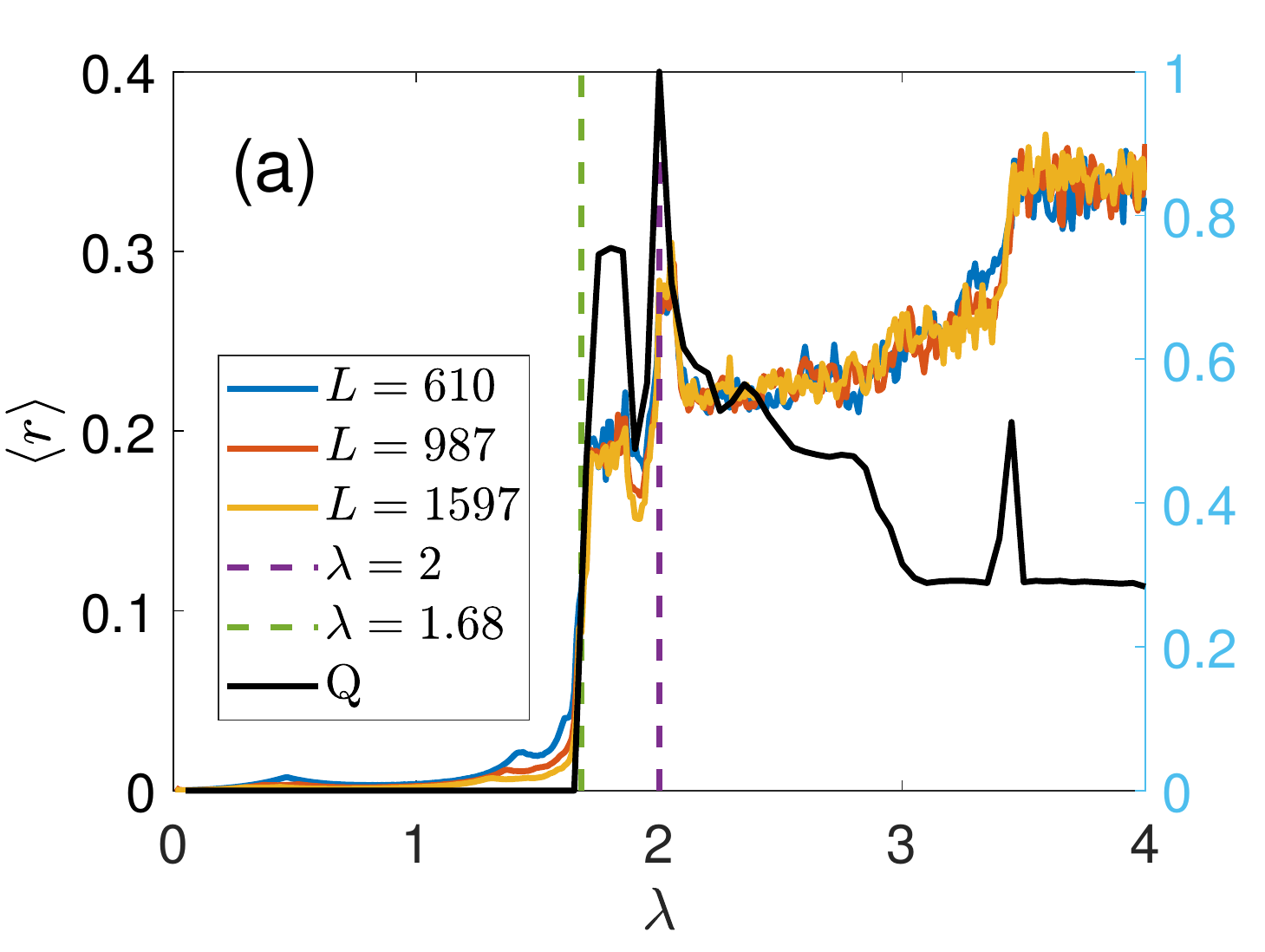}\includegraphics[scale=0.30]{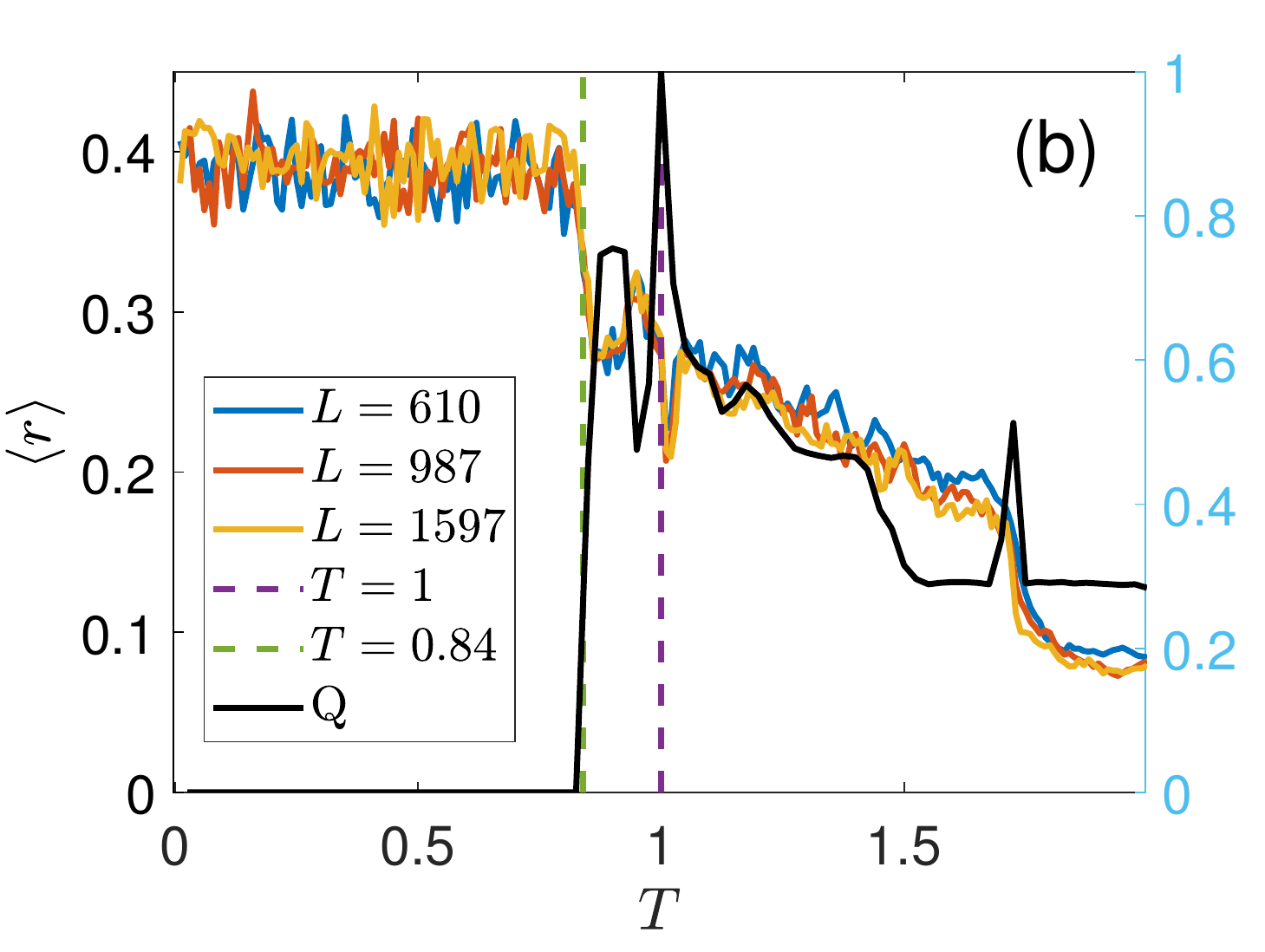}\caption{(a) $\langle r\rangle$ versus $\lambda$ with fixed $T=1$. (b) $\langle r\rangle$ versus $T$ with fixed $\lambda=2$. Black lines (right y axis) depict the fraction of multifractal states for the system with $L=987$ and different parameters. We take 100 samples for each point.\label{fig:transition}}
\end{figure}

Similar to the AA model, we note that the unitary operator fulfills a self-duality relation at $\lambda=2T$ after a dual transformation (see appendix A for details). The existence of a duality mapping suggests that there is a one-to-one
correspondence for parameters which are symmetric about $\lambda=2T$, for
example $T=0.8,\lambda=3$ and $T=1.5,\lambda=1.6$. When across the self-duality point, there
is a sharp transition from localized (extended) to extended (localized) states.
As we discussed above, there are multifractal states in the low frequency
region which can be detected by analyzing the spectrum and eigenvectors.
The transition from extended to multifractal or localized to
multifractal cannot be predicted by the self-duality relation. In
order to study the behavior in the region that the multifractal states
begin to appear, we calculate the $\langle r\rangle$ along the line
$T=1$ and make finite size analysis. As shown in Fig.\ref{fig:transition}, there is a sharp transition when $\lambda\approx1.68$
and this value is smaller than the self-dual point $\lambda/T=2$. When
we increase the system size, we find that the transition of $\langle r\rangle$
around the transition point becomes more and more sharper, which shows a
signature of transition instead of crossover. Fixing the strength of quasiperiodic potential $\lambda=2$ and tuning
the period $T$,  we observe that there is also a sharp change of $\langle r\rangle$
around $T\approx0.8$ as shown in Fig.\ref{fig:transition}.
It is worth  pointing out that a sharp change around the self-duality point $\lambda/T=2$ is always observed in Fig.\ref{fig:transition}(a) and (b).
Such a change is induced by the change of extended (localized) to localized (extended) states in the middle region when across the self-duality point.

Besides, we define a quantity  to describe the fraction of multifractal states:
$$ Q = n_{mul}/n_{all}, $$
where $ n_{mul}$ is the number of multifractal states and $ n_{all}$ is the number of all the eigenstates. We show the change of $Q$ versus $\lambda$ and $T$ for the system with $L=987$ in Fig.\ref{fig:transition}(a) and Fig.\ref{fig:transition}(b), respectively. Below the first transition point, $Q=0$. When the parameters satisfy $\lambda=2T$, all the eigenstates are multifractal and $Q=1$.  The  sharp change of $\langle r \rangle$ has a one-to-one correspondence to the change of $Q$. When $\langle r\rangle$ goes to zero or Poisson value, the fraction of multifractal states approaches to zero, indicating completely delocalized or localized bands, respectively.

In general, multifractal eigenstates or mobility edges can lead to exotic dynamical behaviors. Next we study the expansion dynamics of wavepacket in the region with multifractal states and try to extract the multifractal
exponents from the dynamical behavior. We label the center of the lattice as $j_{0}$ and choose the initial state localized at site of $j_{0}$. The time evolution of an initial state can be expanded by the eigenstates of $U(t)$:
\[
|\psi(t)\rangle=\sum_{m}e^{-i\theta_{m}t}\langle\psi_{m}|\psi(0)\rangle|\psi_{m}\rangle = \sum_{j} C_{j}(t) | j \rangle,
\]
with $C_{j}(t) = \sum_m C^{(m)*}_{j_0}C^{(m)}_{j}e^{-i \theta_m t}$, where $C^{(m)}_{j}=\langle j| \psi_m \rangle$.
Here we focus on the long-time survival probability $P(r)$ defined as
\begin{equation}
P(r)=\sum_{|j-j_{0}|\leq r/2}|C_{j}(t\rightarrow\infty)|^{2},
\end{equation}
which  is the
probability of finding the particle in sites within the region $\left(-r/2,r/2\right)$
after a long time evolution. $P(r)$ is proportional
to $(r/L)^{\widetilde{D_{2}}}$, where $\widetilde{D_{2}}$ is the
generalized dimension of the spectral measure \cite{xiaolong Deng/PRL,XuZH}. For one-dimensional
systems, the dimension of eigenstates fulfills $D_{2}=\widetilde{D_{2}}$\cite{XuZH,Huckestein}.
It is obvious that $D_{2}=0$ in the localized region and $D_{2}=1$ in the extended region.

\begin{figure}
\centering{}\includegraphics[scale=0.30]{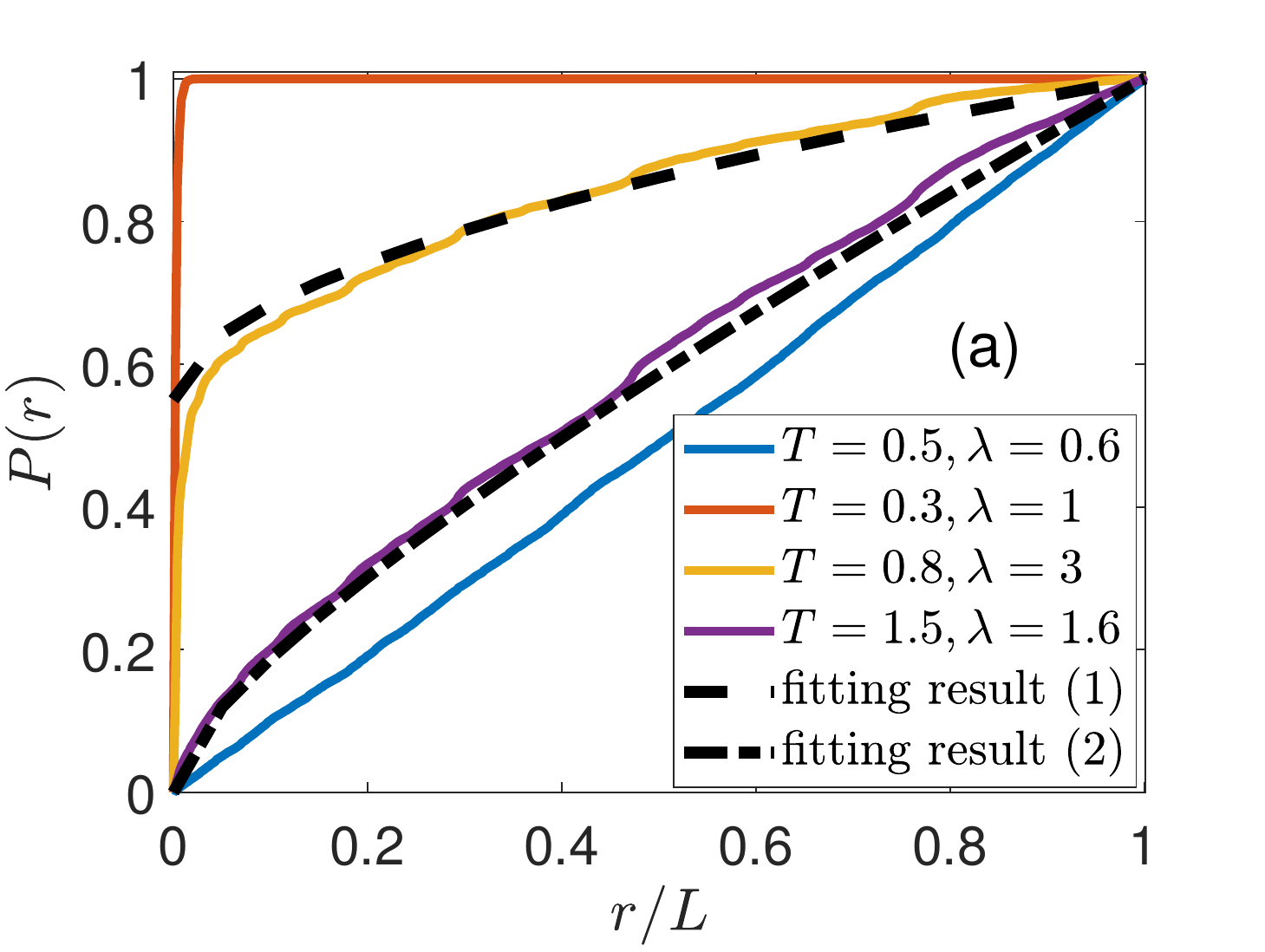}\includegraphics[scale=0.20]{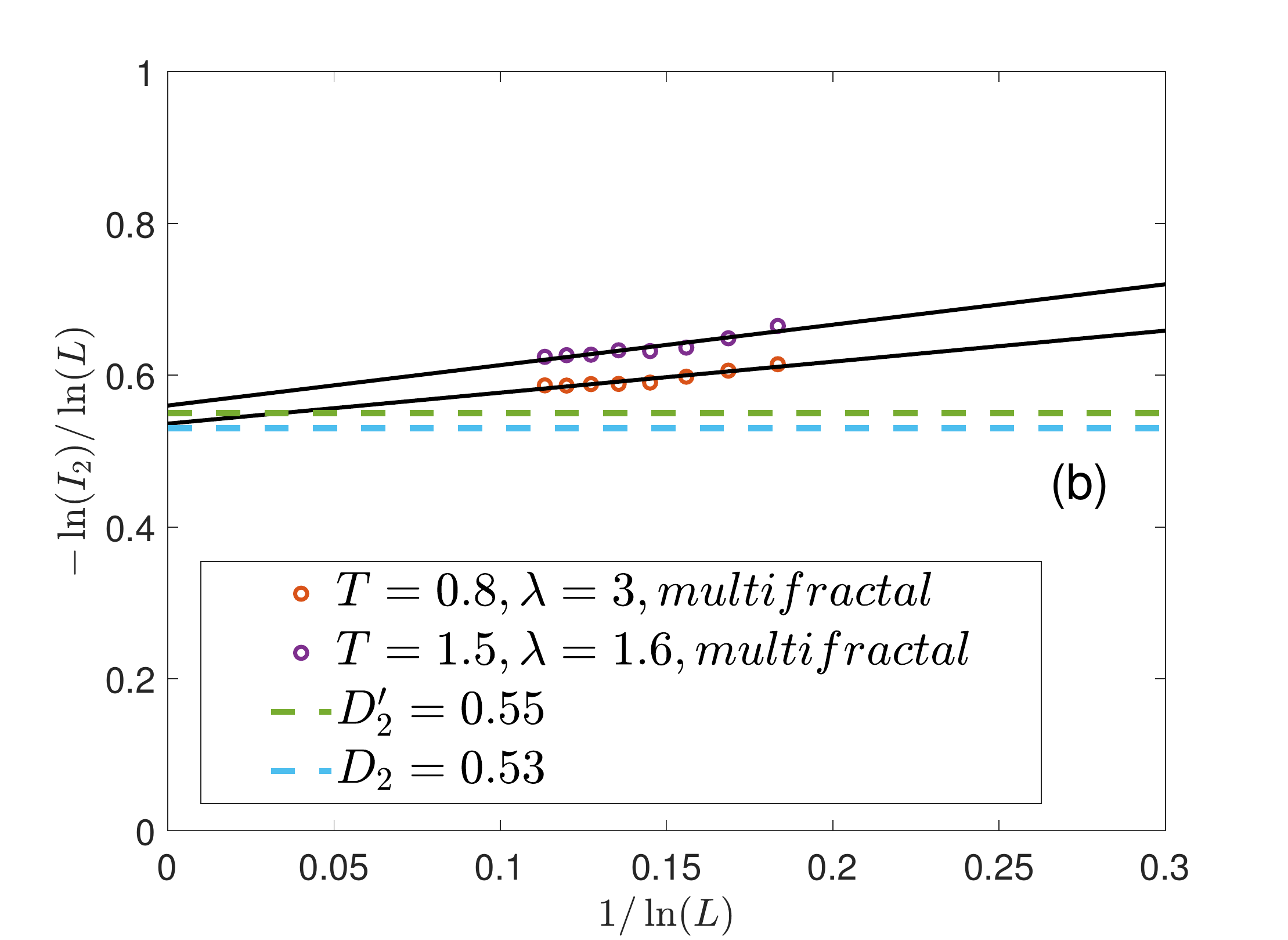}\caption{(a) Long-time survival probability ($t=10^{7}$) with different parameters.
Fitting result (1): we use Eq.(\ref{fitting}) to fit the curve of $T=0.8,~\lambda=3$ with the fitting parameters $c_0=0.55$ and $D_2=0.53$. Fitting result (2):  for the curve of $T=1.5,~\lambda=1.6$, we get $c_0=0.55$ from its dual model. Then we use Eq.(\ref{fitting2}) to fit the curve and get $D_2'=0.55$.
(b) Finite size analysis for multifractal states and we take an average for all the multifractal states with fixed parameters. We choose
$L=987$ and take 1000 samples.  \label{fig:Pr}}
\end{figure}

Consider the case with localized-to-multifractal edge, for which the eigenstates
are either localized or multifractal. While the wavepacket does not expand
in the localized region, the multifractal states play an important role in the
expansion of wavepacket. As shown in Fig.\ref{fig:Pr}, $P(r)$ shows quite different behaviors in the localized region ($T=0.3,\lambda=1$), extended region ($T=0.5,\lambda=0.6$) and region with localized-to-multifractal edge ($T=0.8,\lambda=3$).
In the localized region, all eigenstates are localized, and $P(r)$
grows to 1 rapidly because the wavefunction is mainly distributed
at the initial position. In the extended region, all the eigenstates are extended. $P(r)$ grows uniformly and the wavefunction is distributed in space
uniformly. In the region with localized-to-multifractal edge,
$P(r)$ increase with $r$ but with a nonzero value at $r=0$.
Due to the existence of some localized states, a part of the wavefunction remains at the
initial position. As the increase of $P(r)$ is entirely determined
by the multifractal states, we can extract the average multifractal
exponent by
\begin{equation}\label{fitting}
\ln(P(r)-c_{0})\approx D_{2}\ln(r/L)+\ln(1-c_0),
\end{equation}
where $c_{0}$ is a constant and depends on the proportion of localized states in all the eigenstates \cite{xiaolong Deng/PRL}. $D_{2}$ is determined by the slope of $\ln(P(r)-c_{0})-\ln(r/L)$ line, which gives rise to $D_{2}\approx0.53$.
Because all eigenstates contribute to the time evolution, the multifractal exponents extracted by the wavepacket dynamics should be an average for all mutlifractal states. In Fig.\ref{fig:Pr}(b), we
compare the multifractal exponent extracted from the wavepacket dynamics with
the result from the finite size analysis, which also approaches $0.53$ in the limit $L \rightarrow \infty$.
For the case with extended-to-multifractal edge ($T=1.5,\lambda=1.6$), since both the extended and multifractal
eigenstates attribute to the expansion of the wavepacket, it is hard to read out the multifractal exponent directly from $P(r)$. However, we note that one may roughly estimate the average multifractal exponent $D_2'$ by using the duality property and we get the result $D_2'=0.55$ which is consistent with the result from the finite-size analysis (see the appendix B for details).

\section{Many-body localization in the interacting kicked AA model}
\begin{figure}
\begin{centering}
\includegraphics[scale=0.32]{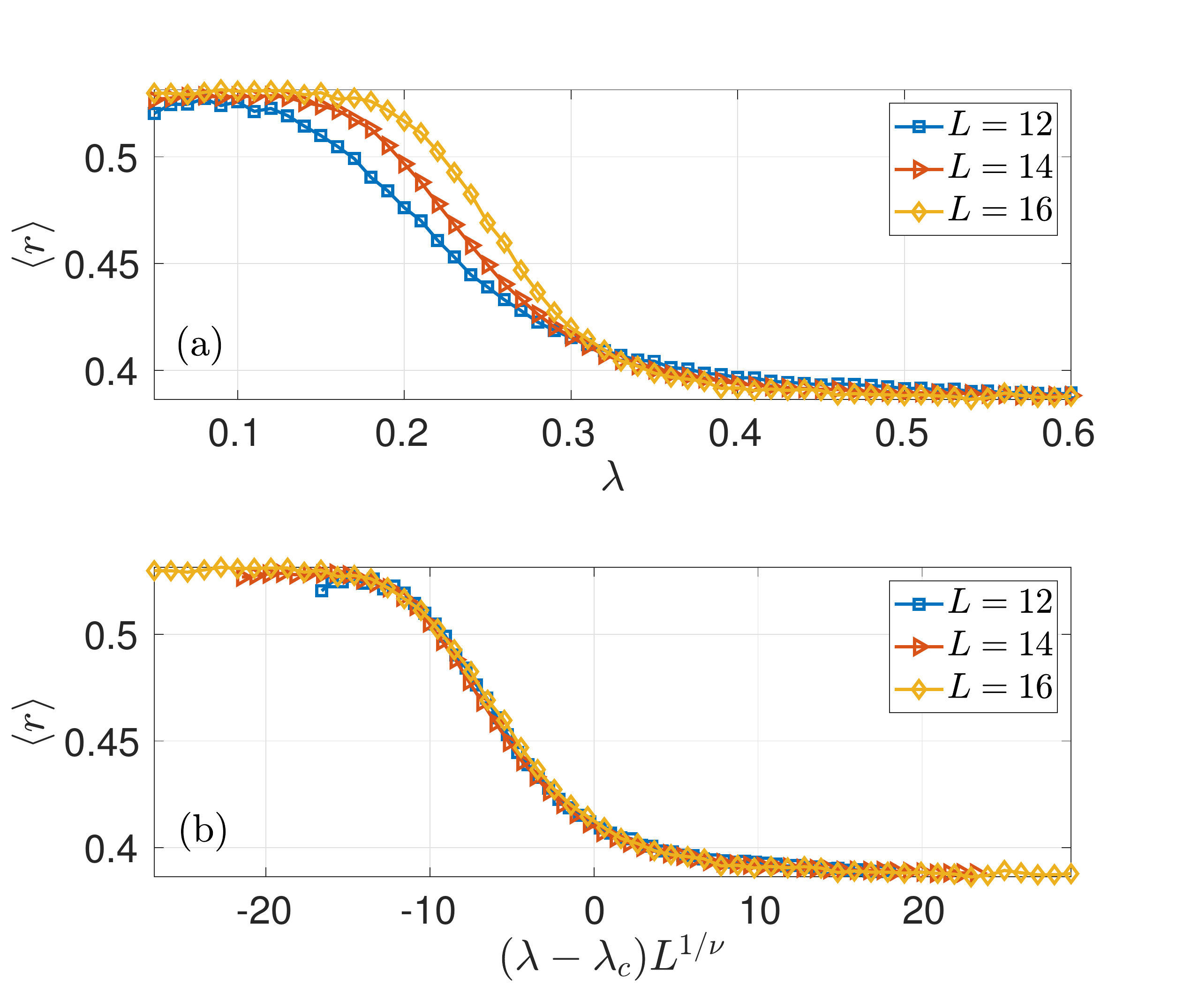}
\par\end{centering}
\caption{(a) The average ratio $\langle r\rangle$ versus $\lambda$ for systems with $T=0.1$ and different sizes. (b) $\langle r\rangle$ versus scaled $\lambda$ with different sizes collapse into a single curve. Each
data point is averaged over 2000 quasiperiodic disorder realizations for $L=12$
and  $L=14$, and 50 realizations for
$L=16$. \label{fig:Finite-size-critical-scaling}}
\end{figure}

Now we study the interacting system with finite $V$ and consider the half-filling case with $N_{f}/L=1/2$, where $N_f$ is the particle number. In the high-frequency limit, we expect the interaction to induce MBL when the strength of quasiperiodic potential exceeds a critical value.  The transition from a dynamical ergodic
phase to MBL phase can also be captured by the average level-spacing ratio $\langle r\rangle$ for the quasienergy spectrum.
In Fig.\ref{fig:Finite-size-critical-scaling} (a), we plot the average
energy level-spacing ratio with a fixed $T=0.1$ versus $\lambda$ for the system with $V=1$ and various system sizes.
We find the level-spacing ratio changes from about $0.52 \pm 0.1$ to $0.39$ when $\lambda$ increases.
The curves with different $L$ intersect at the same point $\lambda_c \approx  0.31$.
By plotting $\langle r\rangle$ versus the scaled potential strength $(\lambda- \lambda_c)/L^{1/\nu}$
for different system sizes, we find that all curves collapse into a single one, as shown in Fig.\ref{fig:Finite-size-critical-scaling} (b).  The finite size analysis gives the transition
point and the critical index as $\lambda_{c}  \approx0.3138$ and $\nu  \approx0.6$ (see appendix C for more details). In the large size limit, it then follows that $\langle r\rangle\approx0.39$ for $\lambda>\lambda_c$ and  $\langle r\rangle\approx0.53$ for $\lambda<\lambda_c$. Our numerical results confirm that quasienergy spectrum statistics follows  a Poisson distribution in the MBL phase and a circular orthogonal ensemble (COE) in the ergodic phase \citep{Abanin MBL Floquet 2015,Abanin Thouless Energy Floquet,meng cheng energy level statistics}.

\begin{figure}
\begin{centering}
\includegraphics[scale=0.24]{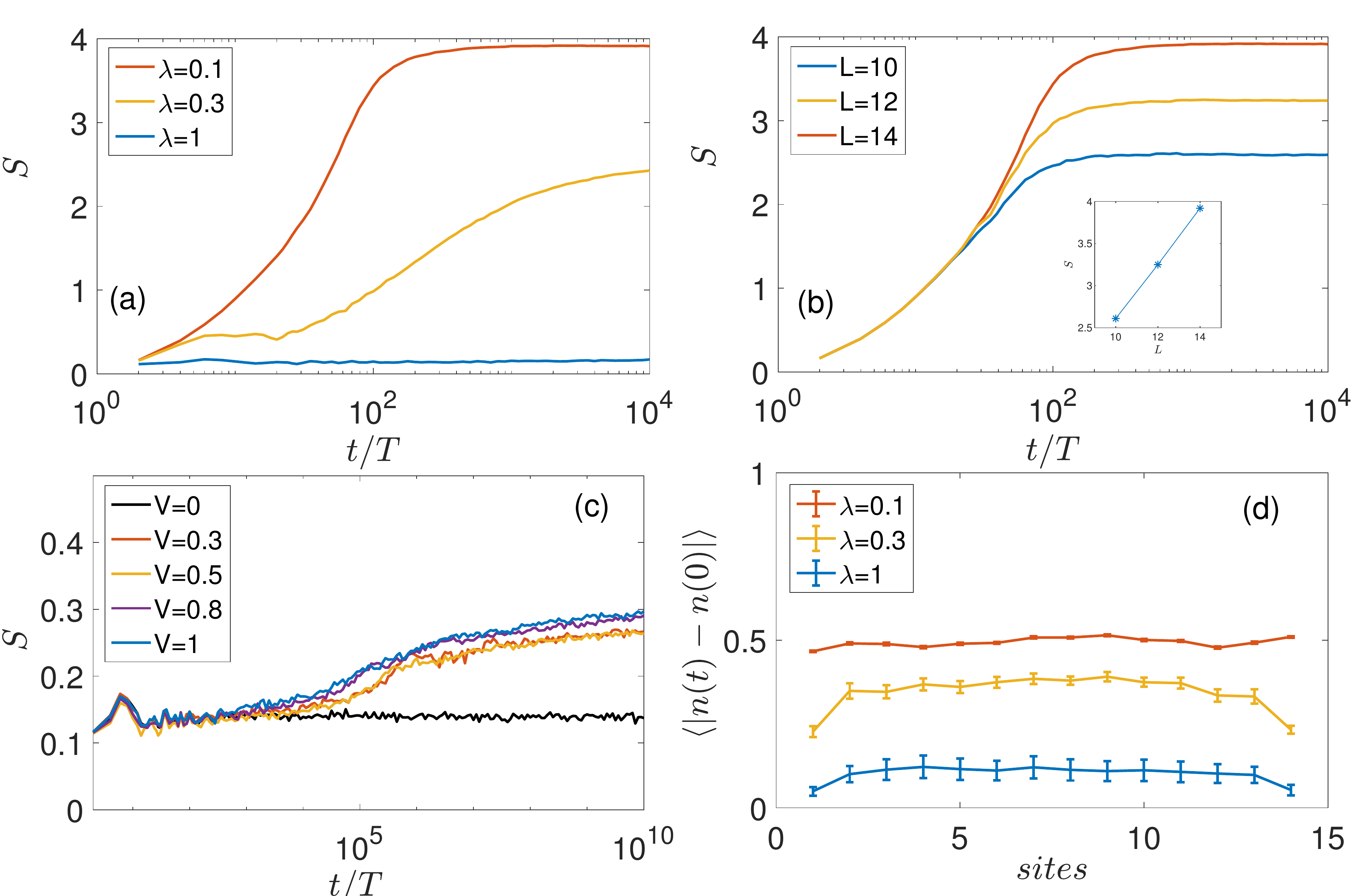}
\par\end{centering}
\centering{}\caption{Dynamical behavior in the kicked interacting AA model with kicked period $T=0.1$. (a) Entanglement entropy growth with  $L=14$ and different quasiperiodic strength $\lambda=0.1$, $0.3$ and $1$. (b) Entanglement entropy growth in the ergodic region with $\lambda=0.1$ and different system sizes
\label{fig:Entanglement-entropy}. The insert in (b) shows the saturation values of the entanglement entropy with different system sizes. (c) The evolution of entanglement entropy in the MBL region with $\lambda=1$ and different interaction strength. (d) Density distribution after $10^4$ kicked periods for system with L=14 and different $\lambda$\label{fig:difference-of-density}. In our calculation, we choose 1000 samples for $L=10$, $12$ and 500 samples for $L=14$.}
\end{figure}

The entanglement entropy is another important parameter to distinguish the
ergodic phase and  MBL phase. The entanglement entropy of the system's eigenstate shows distinct
behavior in different phases:it follows a volume
law in the ergodic phase yet an area law in the MBL phase \citep{MBL-EE2,MBL-EE1}. 
The growth of entanglement entropy  with time in the kicked quasiperiodic lattice is also expected to show different dynamical behaviors in the ergodic and MBL phase.  We
choose the initial state as $|10\dots10\rangle$ with all the odd sites being occupied and all
the even sites empty. In our calculation, we act $U(T)^N$ on the initial state $| \psi(0) \rangle$ to get the finial state $| \psi(NT) \rangle$. In order to calculate
the growth of entanglement entropy, we divide the system into two
parts A and B with the same length and take the trace of subsystem
B to get the reduced density matrix $\rho_{A}$. The entanglement
entropy can be written as
\[
S=-\textrm{tr}\text{(\ensuremath{\rho_{A}\ln\rho_{A}})} .
\]
In Fig.\ref{fig:Entanglement-entropy}a, we display the entanglement entropy growth for $\lambda=0.1$, $0.3$ and $1$, which exhibits distinct behaviors in different phases. In the
ergodic phase with $\lambda=0.1$, the entanglement entropy increases with time and approaches a saturation value (about $3.916$ at $t=10^4T$).
As shown in Fig.\ref{fig:Entanglement-entropy}(b) for systems with $\lambda=0.1$ and different system sizes, the saturation value of the entanglement entropy displays a linear increase with $L$ in the ergodic phase and fulfills the volume law \citep{page entanglement_entropy}. In contrast, the entanglement entropy in the localized phase takes a small value and is not sensitive to the system size. In Fig.\ref{fig:difference-of-density}(c),  we display the long-time behavior of entanglement entropy in MBL phase for different interaction strengthes. It is shown that the entanglement entropy for the interacting systems grows slowly after a long time evolution, whereas the entanglement entropy for the non-interacting system keeps almost unchanged.  Although the entanglement entropy in the MBL phase keeps growing, it is always much smaller than that in the ergodic phase \cite{Moore}.

Further, the dynamics of the system can be intuitively illustrated through the evolution of density distributions.  In Fig.\ref{fig:difference-of-density}(d), we display
the change of real space density distribution
$
\left\langle \left|n_{i}(t)-n_{i}(0)\right|\right\rangle
$
for various $\lambda$, where $n_{i}(t)$ is the time-dependent local density at site
$i$, $\left\langle \dots\right\rangle $ means
sample averages over different phase $\phi$, and $t=NT$ with $N=10^4$ kicked periods. In the ergodic region with $\lambda=0.1$,
$n_{i}(t)$ tends to $0.5$ and it means all the particles are evenly
distributed on the sites after a long time evolution, whereas in the localized
region  with $\lambda=1$ the change of density distribution is small which means the
system retains the initial state information.

\begin{figure}
\centering{}\includegraphics[scale=0.30]{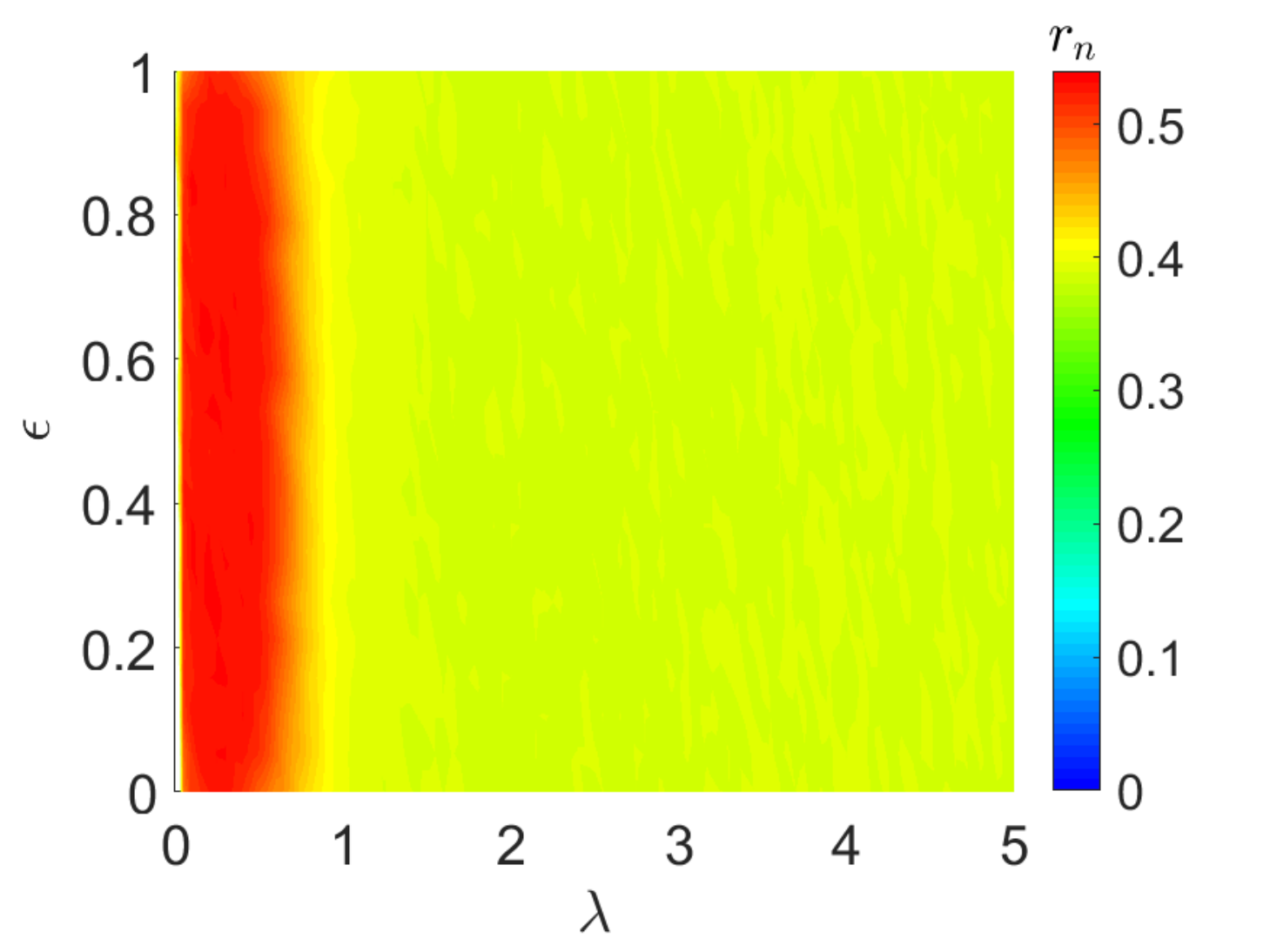}\includegraphics[scale=0.30]{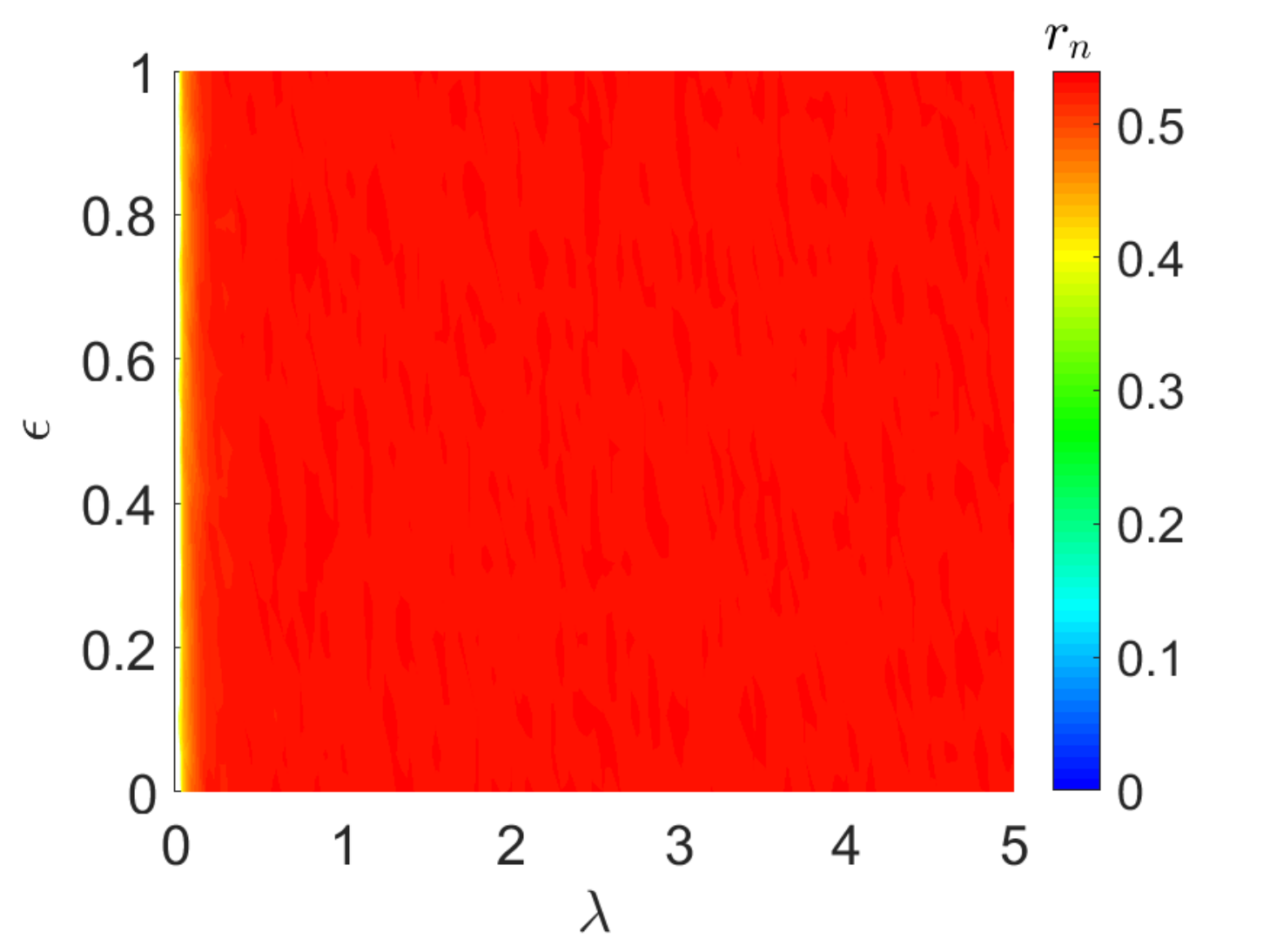}\caption{left: Energy-resolved spectral statistics for $T=0.3$. right: Energy-resolved
spectral statistics for $T=1.5$. We choose $L=14$ and take 1000
samples.\label{fig:left-:energy-resolved-spectral}}
\end{figure}

In the non-interacting case, we have shown the existence of multifractal states and extended/localized-to-multifractal edges in the low-frequency region. Now we study the fate of multifractal states and check whether extended/localized-to-multifractal edges survive in the interacting case.
To this end, we plot the energy-resolved spectral statistics in the parameter space spanned by $\lambda$ and $\epsilon$ for $V=1$, $T=0.3$ and $T=1.5$
in Fig.\ref{fig:left-:energy-resolved-spectral}, where $\epsilon$ is defined as
\[
\epsilon=\frac{\theta_n-\theta_{min}}{\theta_{max}-\theta_{min}},
\]
which labels the place of the quasi-energy $\theta_n$ lying in the quasi-energy density spectrum.
We note that the energy-resolved spectral statistics have been used to characterize the energy-resolved MBL  in Ref.\citep{energy-resolved}.
In the high frequency
region with $T=0.3$, it is shown that a transition from ergodic to MBL phase occurs when we
increase $\lambda$. On the other hand, in the low frequency region with $T=1.5$, we do not observe such a transition and the system is always in the ergodic phase. Our result shows no signature for the existence of extended/localized-to-multifractal edges in the interacting system.
\begin{figure}
\begin{centering}
\includegraphics[scale=0.6]{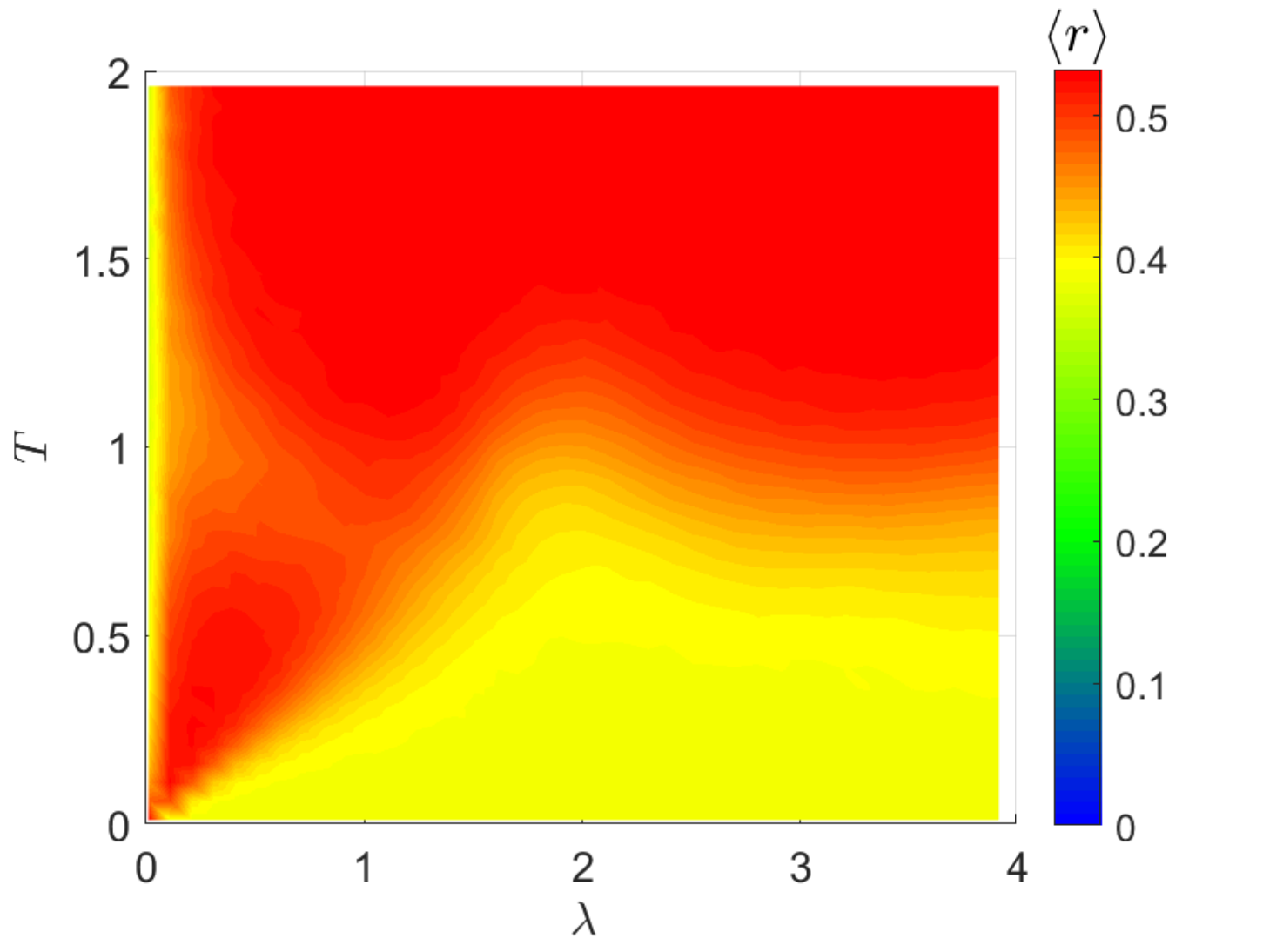}
\par\end{centering}
\caption{The average ratio  $\langle r\rangle$ in the parameter
space spanned by $\lambda$ and T for $V=1$ and $L=12$. We take 1000 samples in our simulation.\label{fig:phasediagram}}
\end{figure}

\begin{figure}
\begin{centering}
\includegraphics[scale=0.6]{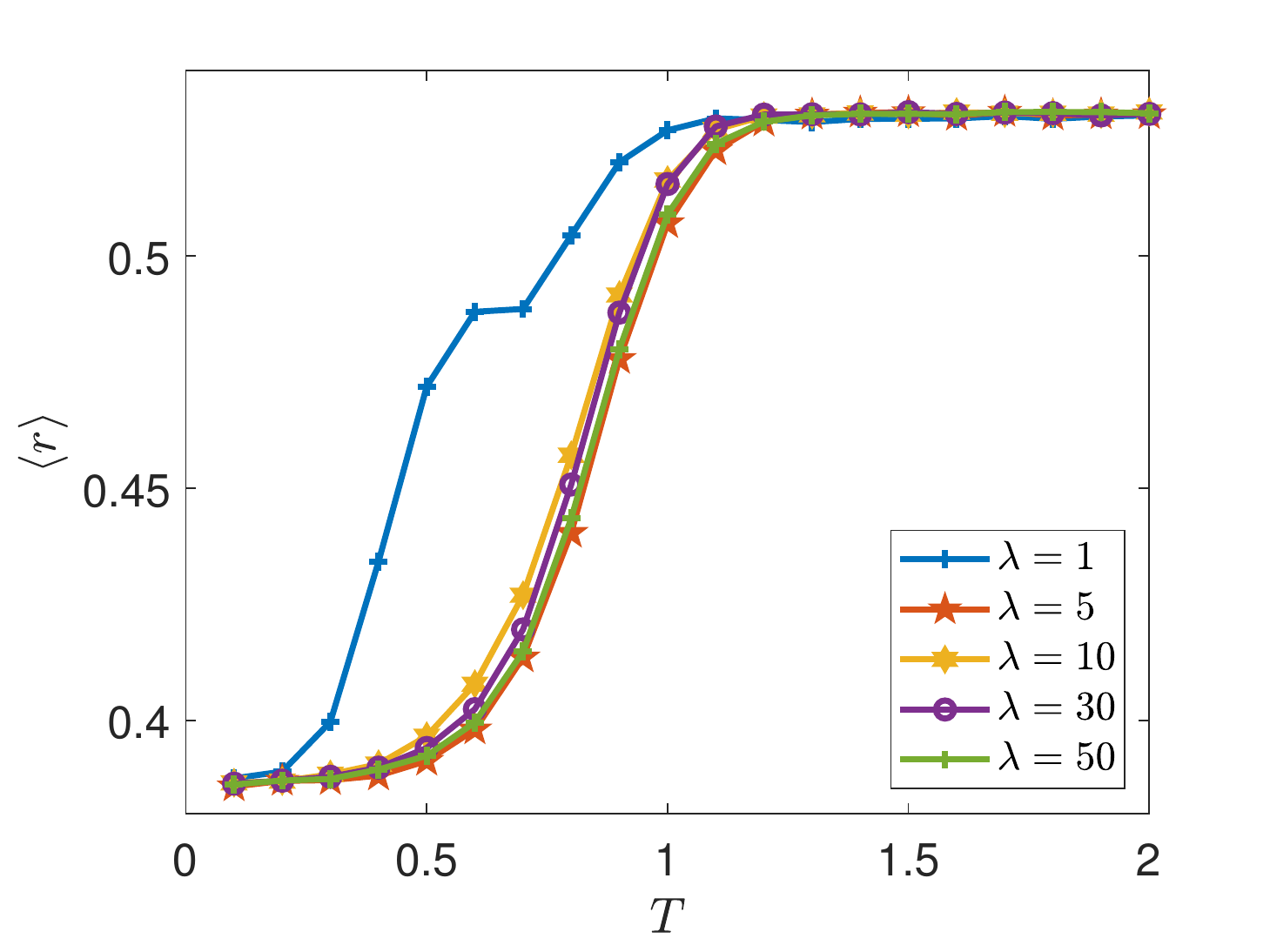}
\par\end{centering}
\caption{The average ratio $\left\langle r\right\rangle $ versus  T with different quasiperiodic
strength $\lambda$. \label{figr-T} The system size is L=14 and we take the average over 500 samples.}
\end{figure}

In Fig.\ref{fig:phasediagram}, we show the average level-spacing ratio  $\langle r\rangle$ in the parameter
space spanned by $\lambda$ and T for a system of $L=12$ with the interaction strength $V=1$. In comparison with
the non-interacting case, we find that the interaction term can lead to the appearance of the MBL phases in the regime with all the eigenstates being localized in the non-interacting limit.
However, in the low-frequency region with the corresponding eigenstates in the non-interacting limit being either extended or
multifractal states, adding an interaction term leads to the thermalization
of the system, characterized by $\left\langle r\right\rangle \approx 0.53$. No signature of MBL is observed by further increasing $\lambda$.
In Fig.\ref{figr-T}, we display $\left\langle r\right\rangle$ versus $T$ for system with the interaction strength $V=1$ and different strength of quasiperiodic potential $\lambda$ (discussion on effect of the interaction strength can be found in the appendix D). In the low-frequency region, it is shown that the ratio $\left\langle r\right\rangle$  increases with the increase of $T$ for various $\lambda$ and approaches $0.53$. Our results show that the MBL vanishes when the systems enter the low-frequency region, and the systems are ergodic even for a very large  $\lambda$. The absence of MBL in the low-frequency region is related to the emergence of  localized-to-multifractal edges in the quasienergy spectrum of non-interacting kicked AA model discussed in the previous section. The presence
of a localized-to-multifractal edge means that both localized and multifractal single-particle orbitals are present and their interplay to the interaction may induce the absence of MBL. Although MBL can occur in the static interacting systems with single-particle mobility edge \cite{MBL-inc1,MBL-inc2}, it has been shown that the presence of a mobility edge anywhere in the spectrum is enough to induce delocalization for any driving strength and frequency \cite{F_MBL1}. Our model, however, provides a different scenario in which either the presence or absence of MBL is possible by tuning the driving frequency.

\section{Summary}

In summary, we have studied the phenomenon of dynamical localization and many-body localization as well as their breakdown in the periodically kicked
AA model. By analyzing the quasienergy spectrum statistics in the non-interacting limit, we have verified the existence of dynamical localization transition in the high-frequency region, which is characterized by an abrupt change of average quasienergy level-spacing ratio $\langle r \rangle$ across the self-dual point $\lambda/T=2$. On the other hand, the spectrum statistics becomes intricate in the low-frequency region due to the emergence of the extended/localized-to-multifractal edges in the quasienergy spectrum, which separate the multifractal states from the localized (extended) states.
We also find that there is a sharp transition when the multifractal states occur in the low frequency region. Furthermore, we discuss the dynamical behavior in different regions and extract the multifractal exponent from the long-time survival probability.
For the interacting  periodically  kicked AA model, we have found the occurrence of a transition from the ergodic phase to the MBL phase in the high-frequency
region,  when the quasiperiodic potential strength exceeds a critical value.
The transition point and the
critical exponent of the ergodic-MBL transition are obtained by a finite-size scaling analysis. We also calculate
the time evolution of entanglement entropy and the density
distribution to confirm the existence of the MBL phase.
We find that the interaction can lead to the thermalization of the system when there are multifractal states in the non-interacting limit,
and demonstrate that the MBL phase vanishes even for strong quasiperiodic potential. Our results show that the interplay of quasiperiodic disorder, driven period, and interaction can lead to rich dynamical phenomena in the periodically kicked AA model.

Note added: Recently, we became aware of the experimental realization of the kicked AA quasiperiodic model studied in the present work and the study of multifractality in a related parallel experimental work\cite{exp-kickedAA}.

\begin{acknowledgments}
The work is supported by National Key
Research and Development Program of China (Grant No.
2021YFA1402104), the NSFC under Grants No.12174436, No.11974413
and No.T2121001 and the Strategic Priority Research Program of Chinese Academy of Sciences under Grant No. XDB33000000.
\end{acknowledgments}

\appendix

\section{Dual transformation for the Floquet unitary propagator of kicked AA model.\label{sec:The-self-dual}}
The phase diagram of kicked AA model displays a symmetrical structure about the diagonal line $\lambda=2T$ and this is
due to the existence of a duality mapping for the kicked AA model.
The Floquet unitary propagator is given by
\[
U(T)=e^{-i\sum_{j=1}^{L}\left(c_{j}^{\dagger}c_{j+1}+h.c.\right)T}e^{-i\lambda\sum_{j}^{L}\cos\left(2\pi\alpha j+\phi\right)\hat{n}_{i}}.
\]
For convenience, we fix $\phi=0$ and take a Fourier transform $c_{j}=\sum_{k}c_{k}e^{i2\pi\alpha kj}$. It follows
\begin{align*}
& U\text{(T,0)} \\
= & e^{-i\sum_{j=1}^{L}(c_{j}^{\dagger}c_{j+1}+h.c.)T}e^{-i\lambda\sum_{j=1}^{L}\cos\left(2\pi\alpha j\right)c_{j}^{\dagger}c_{j}}\\
= & \exp(-i\sum_{j=1}^{L}\sum_{k=1}^{L}\sum_{k'=1}^{L}(c_{k}^{\dagger}c_{k'}e^{i2\pi\alpha k'j-i2\pi\alpha k(j+1)}+h.c.)T)\\
\times & \exp(-i\lambda\sum_{j=1}^{L}\sum_{k=1}^{L}\sum_{k'=1}^{L}\frac{1}{2}(e^{i2\pi\alpha j}+e^{-i2\pi\alpha j})c_{k}^{\dagger}c_{k'}e^{i2\pi \alpha (k-k')j})\\
\approx & e^{-i\sum_{k=1}^{L}\left(2\cos(2\pi\alpha k)\right)c_{k}^{\dagger}c_{k}T}e^{-i\lambda\sum_{k=1}^{L}\frac{1}{2}(c_{k}^{\dagger}c_{k+1}+c_{k+1}^{\dagger}c_{k})} .
\end{align*}
In the last step, we use ``$\approx$'' instead of ``$=$'' because we use an approximation $\sum_{j=1}^L e^{i 2\pi\alpha (k-k')j}\approx\delta(k-k')$, which holds true exactly only in the limit of $L \rightarrow \infty$ or by taking $\alpha=M/L$ to approximately represent an irrational number with $M$ being coprime to $L$.
Using $X$ to represent the transformation, we get the following relation:
\[
X^{-1}U(T,\lambda)X\approx U^{\dagger}(-\frac{\lambda}{2},-2T).
\]
According the properties of Floquet operator $U^{\dagger}(T)=U^{-1}(T)$, we can get
\[
X^{-1}U(T,\lambda)X\approx U^{-1}(\frac{\lambda}{2},-2T).
\]
After the transformation, we can see that the hopping term in real space corresponds to the on-site potential term in momentum space, and the on-site potential term in real space corresponds to the hopping term in momentum space. It is clear that $\lambda=2T$ is the self-duality point. We note that the duality mapping is also discussed in Ref.\cite{exp-kickedAA}.

\section{Multifractal exponent in the case with extended-to-multifractal edge \label{sec:extended-multifractal exponent}}
For the case with extended-to-multifractal edge, both the extended and multifractal
eigenstates attribute to the expansion of the wavepacket. In order to extract the average multifractal exponent, we need
to eliminate the effect of the extended states. According to the duality
properties, the position of the extended-to-multifractal edge is the
same as the position of the localized-to-multifractal edge of its dual model if the
parameters satisfy the dual mapping relation. Extended eigenstates lead
to a linear increase of the long-time survival probability $P(r)$. We can
divide the eigenstates into two parts: all the eigenvalues are
extended in the first part  and all the eigenvalues are
multifractal in the second part. We can estimate the number
of extended states by the number of localized states in its dual model.
We assume that the proportion of extended states in all eigenstates
is $p$ and we define a modified long-time survival probability:
\begin{equation}
P_{1}(r)=P(r)-p\times\frac{r}{L},
\end{equation}
where the second term is used to eliminate the effect of the extended
states. We can extract the average multifractal exponent by
\begin{equation}\label{fitting2}
\ln(P_{1}(r))\approx D_{2}'\ln(r/L)+\ln(1-p),
\end{equation}
where $D_{2}'$ is determined by the slope of $\ln(P_{1}(r))-\ln(r/L)$ line and $p=c_0$ with $c_0$ determined from its dual model.

\section{Finite-size scaling analysis \label{sec:finite-size scaling analysis}}
As we know, the ratio $\langle r\rangle$ changes from 0.53 to 0.39
when the system undergoes a transition from ergodic phase to MBL phase. We perform
a finite-size scaling analysis for $\langle r\rangle=f\left[(\lambda-\lambda_{c})L^{1/\nu}\right]$
with a fixed kicked period $T=0.1$ and interaction strength $V=1$.
Here $\lambda_{c}$ denotes the transition point from the ergodic phase
to MBL phase and $\nu$ is the associated critical exponent. We fit
numerical data for the region close to the phase transition by Taylor
expanding the scaling function $f\left[(\lambda-\lambda_{c})L^{1/\nu}\right]$
and the scaling variables
\[
\langle r\rangle\approx f^{(0)}+f^{(1)}(\lambda-\lambda_{c})L^{1/\nu}+\dots .
\]
By performing a non-linear least squares fitting, we find the best fitting is
$\lambda_{c}\approx0.3138$ and $\nu\approx0.6$.

\section{Effect of the interaction strength \label{sec:the effect of the interaction strength}}
\begin{figure}
\centering{}\includegraphics[scale=0.6]{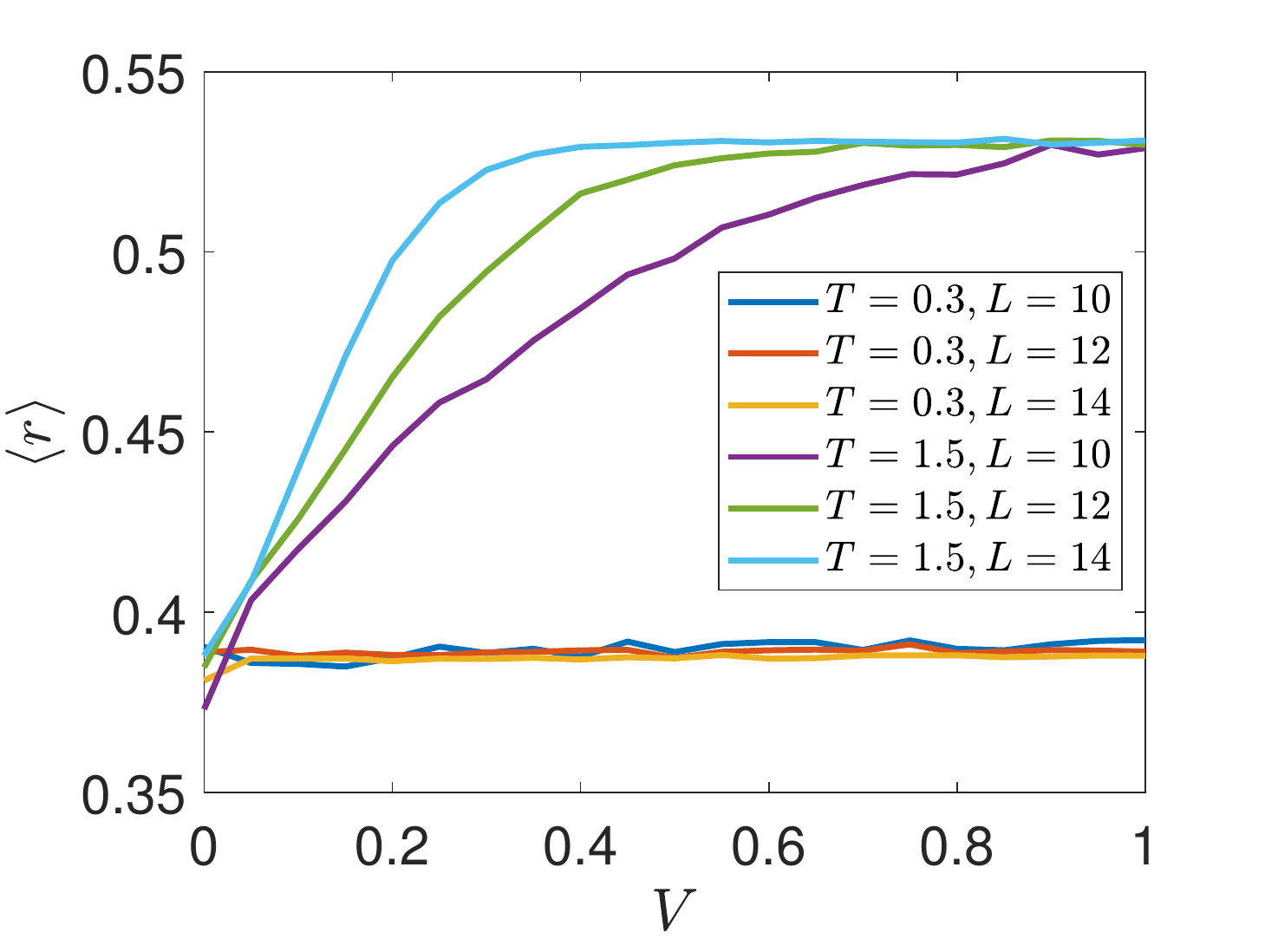}\caption{$\langle r\rangle$ in different kick periods and system sizes. We
choose $\lambda=3.5$ and take 1000 samples for every curve.\label{fig:label}}
\end{figure}

In the discussion of the main text, we choose the interaction strength
as $V=1$. Here we discuss the effect of interaction strength on the
thermalization of the system. The results are shown in Fig.\ref{fig:label}.
We choose $\lambda=3.5$ and $T=0.3$ and $1.5$, respectively. In the non-interacting
case, all eigenstates are localized for $T=0.3$ (case 1) and there
are multifractal-to-localized edges for $T=1.5$ (case 2). As we can
see, for the case 1, $\langle r\rangle$ remains at 0.39 in the MBL region when
we increase the interaction strength and the system size. In contrast, for the case 2, increasing interaction strength can lead to the thermalization
of the system. As we increase the size of the system, smaller interaction
can lead to thermalization of the system.

\end{document}